\documentclass[9pt,twocolumn,twoside]{optica}
\setboolean{shortarticle}{false}
\usepackage{amsmath}
\usepackage{amsfonts}

\newcommand{\real}{\operatorname{Re}}
\newcommand{\imag}{\operatorname{Im}}
\newcommand{\parti}[2]{\frac{\partial #1}{\partial #2}}

\newcommand{\intall}{\int_{-\infty}^{\infty}}
\newcommand{\ket}[1]{|#1\rangle}

\newcommand{\bra}[1]{\langle#1|}

\newcommand{\avg}[1]{\langle#1\rangle}
\newcommand{\Avg}[1]{\left\langle#1\right\rangle}

\newcommand{\bs}[1]{\boldsymbol{#1}}
\newcommand{\abs}[1]{\left|#1\right|}
\newcommand{\bk}[1]{\left(#1\right)}
\newcommand{\Bk}[1]{\left[#1\right]}
\newcommand{\BK}[1]{\left\{#1\right\}}

\newcommand{\trace}{\operatorname{tr}}

\newcommand{\expect}{\mathbb E}

\title{Quantum limits to optical point-source localization}

\author[1,2,*]{Mankei Tsang}
\affil[1]{Department of Electrical and Computer Engineering,
  National University of Singapore, 4 Engineering Drive 3, Singapore
  117583}

\affil[2]{Department of Physics, National University of Singapore,
  2 Science Drive 3, Singapore 117551}

\affil[*]{\url{mankei@nus.edu.sg}}

\dates{Compiled \today}

\ociscodes{(270.5585) Quantum information and processing;
(270.6570) Squeezed states; (100.6640) Superresolution.}

\doi{}

\begin{abstract}
  Motivated by the importance of optical microscopes to science and
  engineering, scientists have pondered for centuries how to improve
  their resolution and the existence of fundamental resolution
  limits. In recent years, a new class of microscopes that overcome a
  long-held belief about the resolution have revolutionized biological
  imaging. Termed ``superresolution'' microscopy, these techniques
  work by accurately locating optical point sources from far field. To
  investigate the fundamental localization limits, here I
  derive quantum lower bounds on the error of locating point sources
  in free space, taking full account of the quantum, nonparaxial, and
  vectoral nature of photons.  These bounds are valid for any
  measurement technique, as long as it obeys quantum mechanics, and
  serve as general no-go theorems for the resolution of microscopes.
  To arrive at analytic results, I focus mainly on the cases of one
  and two classical monochromatic sources with an initial vacuum
  optical state. For one source, a lower bound on the root-mean-square
  position estimation error is on the order of $\lambda_0/\sqrt{N}$,
  where $\lambda_0$ is the free-space wavelength and $N$ is the
  average number of radiated photons. For two sources, owing to the
  statistical effect of nuisance parameters, the error bound diverges
  when their radiated fields overlap significantly. The use of
  squeezed light to enhance further the accuracy of locating one
  classical point source and the localization limits for partially
  coherent sources and single-photon sources are also discussed. The
  presented theory establishes a rigorous quantum statiscal inference
  framework for the study of superresolution microscopy and points to
  the possibility of using quantum techniques for true resolution
  enhancement.
\end{abstract}

\begin{document}
\maketitle
\thispagestyle{fancy}

\section{Introduction}
The resolution limit of optical microscopes is one of the most
important problems in science and engineering. The Abbe-Rayleigh
criterion with respect to the free-space wavelength $\lambda_0$ has
been a widely used resolution limit \cite{born_wolf}, but it is now
well known that the criterion is heuristic in the context of
microscopy and superresolution microscopy is possible. An important
class of superresolution microscopy, including
stimulated-emission-depletion microscopy \cite{hell} and
photoactivated-localization microscopy \cite{betzig}, relies on the
accurate localization of point sources from far field \cite{moerner}.
The localization accuracy, which represents an important measure of
the microscope resolution, is then limited by the statistics of the
optical measurement \cite{bobroff,thompson02,ober,ram}. Prior analyses
of point-source localization accuracy assume classical, scalar, and
paraxial optics with statistics specific to the measurement methods
\cite{bobroff,thompson02,ober,ram}. On a more fundamental level,
however, optics is governed by the quantum theory of electromagnetic
field \cite{mandel}, and the existence of more accurate measurement
methods \cite{sheppard07} and more fundamental quantum limits remains
an open question. For example, the superoscillation phenomenon
\cite{berry_popescu} suggests that superresolution diffraction
patterns can be obtained in the expense of signal power; can it be
exploited to improve the resolution of microscopes
\cite{zheludev08,huang_zheludev,hyvarinen12}?

Using a quantum Cram\'er-Rao bound (QCRB) \cite{helstrom,holevo} and
the full quantum theory of electromagnetic field \cite{mandel}, here I
derive quantum limits to the accuracy of locating point sources. These
quantum resolution limits are more general and fundamental than prior
classical analyses in the sense that they apply to any measurement
method and take full account of the quantum, nonparaxial, and vectoral
nature of photons.  To arrive at analytic results, I focus mainly on
the cases of one and two monochromatic classical sources and an
initial vacuum optical state. The possibility of using squeezed light
to further enhance the accuracy of locating one point source will also
be discussed.  To study partially coherent sources, I model
incoherence using the concept of nuisance parameters, which are
unknown parameters of no primary interest in the context of
statistical inference. Quantum bounds for partially coherent sources
are then derived by introducing a new generalized QCRB that accounts
for nuisance parameters in a special way.

In quantum optics, there has been a substantial literature on quantum
imaging; see, for example,
Refs.~\cite{helstrom,helstrom70,kolobov,fabre,treps,barnett,boto,li08,
  boyd2012,centroid,shin,rozema,glm_imaging,nair_yen,pirandola,perez12,hemmer12,humphreys,schwartz_oron,schwartz13,cui13,monticone,taylor2013},
but most of them assume certain quantum optical states without
considering how they may be generated by objects relevant to
microscopy or consider simply the estimation of mirror
displacement. Helstrom's derivation of the QCRB for one point source
\cite{helstrom,helstrom70} is the most relevant prior work, although
he used the paraxial approximation, did not consider the use of
squeezed light, and studied two sources only in the context of binary
hypothesis testing \cite{helstrom}. There have also been intriguing
claims of superresolution using the nonclassical photon statistics
from single-photon sources
\cite{schwartz_oron,schwartz13,cui13,monticone}, but their protocols
have not been analyzed using statistical inference, so even though
their images appear sharper, the accuracies of their methods in
estimating object parameters remain unclear. To investigate their
claims, I also derive a quantum bound for locating a single-photon
source.

\section{Quantum parameter estimation}
\label{sec_qcrb}
Let the initial quantum state of a system be $\ket{\psi}$.  After
unitary evolution $U(X,T)$ with respect to Hamiltonian $H(X,t)$ as a
function of parameters $X = (X_1,X_2,\dots)$, the quantum system is
measured with outcome $Y$.  The probability distribution of $Y$
according to Born's rule can be expressed as
\cite{helstrom,holevo,wiseman_milburn}
\begin{align}
P(Y|X) &= \trace\Bk{E(Y)U(X,T) \ket{\psi}\bra{\psi} U^\dagger(X,T)},
\end{align}
where $E(Y)$ is the positive operator-valued measure (POVM) that
characterizes the quantum measurement and $\trace$ denotes the
operator trace.  Denote the estimator of $X$ using $Y$ as
$\tilde X(Y)$.  The estimation error matrix is defined as
\begin{align}
\Sigma_{\mu\nu}(X) &\equiv \int dYP(Y|X)
\Bk{\tilde X_\mu(Y)-X_\mu}\Bk{\tilde X_\nu(Y)-X_\nu}.
\label{Sigma}
\end{align}
For unbiased estimators, the classical Cram\'er-Rao bound states that
\cite{vantrees}
\begin{align}
\Sigma(X) &\ge j^{-1}(X),
\end{align}
which means that $\Sigma-j^{-1}$ is positive semidefinite.  $j(X)$ is
the Fisher information matrix given by
\begin{align}
j_{\mu\nu}(X) &\equiv \int dY P(Y|X)\Bk{\parti{}{\theta_\mu}\ln P(Y|X)}
\Bk{\parti{}{\theta_\nu}\ln P(Y|X)}.
\end{align}
The bound has been used, for example, in Refs.~\cite{ober,ram} to
evaluate the point-source localization accuracy for a microscope.

It turns out that, for any POVM and thus any measurement in quantum
mechanics, another lower bound exists in the form of the QCRB
\cite{helstrom,braunstein}:
\begin{align}
\Sigma(X) &\ge j^{-1}(X) \ge J^{-1}(X),
\end{align}
which means that $j^{-1} - J^{-1}$ and $\Sigma-J^{-1}$ are
positive-semidefinite.  $J$ is the quantum Fisher information (QFI)
matrix; it can be obtained by expressing the fidelity
$|\bra{\psi}U^\dagger(X,T)U(X+\delta X,T)\ket{\psi}|^2$ in the
interaction picture \cite{tsang_nair} and expanding it to the second
order of $\delta X$ \cite{paris,pasquale}. The result is
\begin{align}
J_{\mu\nu}(X) &= 4\real \Avg{\Delta g_\mu(X)\Delta g_\nu(X)},
\end{align}
where $\real$ denotes the real part, $\avg{A} \equiv \bra{\psi}A\ket{\psi}$, 
$\Delta A \equiv A-\avg{A}$, and 
\begin{align}
g_\mu(X) &\equiv \frac{1}{\hbar}\int_0^T dt U^\dagger(X,t)
\parti{H(X,t)}{X_\mu} U(X,t)
\end{align}
is the generator of the parameter shift in the Heisenberg picture.
For $M$ trials, the QFI is simply multiplied by $M$, and at least one
component of the QCRB can be attained in an asymptotic $M\to\infty$
sense \cite{fujiwara2006}. If one wishes to consider a new set of
parameters $\theta$ related to the original set $X$ and $X$ can be
expressed as a function of $\theta$, the new QFI matrix is simply
given by
\begin{align}
  J_{ab}'(\theta) =
  \left.\sum_{\mu,\nu}\parti{X_\mu}{\theta_a}J_{\mu\nu}(X)\parti{X_\nu}{\theta_b}
\right|_{X = X(\theta)}.
\end{align}
Various generalizations of the QCRB and alternatives are available
\cite{helstrom,yuen_lax,twc,qzzb,qbzzb}, but the presented version
suffices to illustrate the pertinent physics.  In
Sec.~\ref{sec_qcrb_nuisance}, the QCRB will be generalized to a
Bayesian version that treats nuisance parameters separately and is
used to study partially coherent sources.

\begin{figure}[htbp]
\centerline{\includegraphics[width=0.4\textwidth]{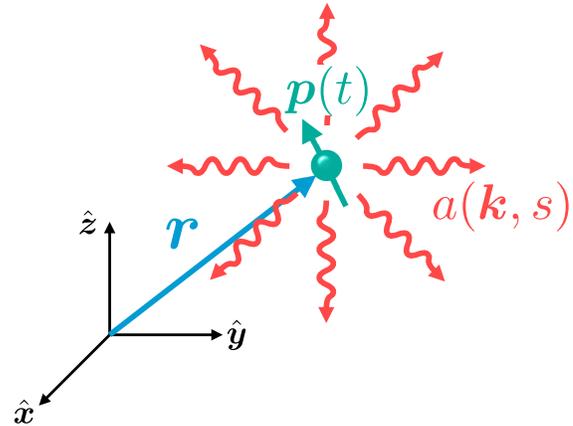}}
\caption{A classical point source with dipole moment
  $\bs p(t)$ radiating in free space. Its position $\bs r$ is
  estimated by measuring the quantum optical field, with $a(\bs k,s)$
  denoting its annihilation operator.}
\label{1source}
\end{figure}

\section{One classical point source}
\label{sec_onesource}
Consider first a classical point source, as depicted in
Fig.~\ref{1source}. The Hamiltonian is \cite{mandel}
\begin{align}
H(\bs r,t) &= H_F + H_I(\bs r,t),
\\
H_F &= \sum_s \int d^3k\, \hbar\omega(\bs k) a^\dagger(\bs k,s) a(\bs k,s),
\\
H_I(\bs r,t) &= -\bs p(t)\cdot \bs E(\bs r),
\\
\bs E(\bs r) &= \sum_s \int d^3k \sqrt{\frac{\hbar\omega}{2(2\pi)^3\epsilon_0}}
\big[i\bs\varepsilon(\bs k,s)a(\bs k,s)e^{i\bs k\cdot\bs r} 
\nonumber\\&\quad
+\textrm{H.c.}\big],
\end{align}
where $\bs k = k_x\hat{\bs x}+k_y\hat{\bs y} + k_y\hat{\bs z}$ is a
wavevector, $(\hat{\bs x},\hat{\bs y},\hat{\bs z})$ denote unit
vectors in the Cartesian coordinate system, $\int d^3k \equiv \intall
dk_x \intall dk_y \intall dk_z$, $s$ is an index for the two
polarizations, $\bs\varepsilon(\bs k,s)$ is a unit polarization
vector, $\omega(\bs k) = c|\bs k|$, $c$ is the speed of light, $\bs
p(t)$ is the c-number dipole moment of the source, $\bs r =x\hat{\bs
  x} + y\hat{\bs y}+z\hat{\bs z}$ is its position, $\epsilon_0$ is the
free-space permittivity, $a(\bs k,s)$ is an annihilation operator
obeying the commutation relation $\Bk{a(\bs k,s),a^\dagger(\bs k',s')}
= \delta_{ss'}\delta^3(\bs k-\bs k')$, and $\textrm{H.c.}$ denotes the
Hermitian conjugate. Since $\bs p(t)$ is a c-number, $H_I$ implements
a field displacement operation \cite{mandel}.  The Heisenberg picture
of $a(\bs k,s)$ is
\begin{align}
a(\bs k,s,t) &\equiv U^\dagger(X,t)a(\bs k,s)U(X,t)
\\
&= e^{-i\omega t}\Bk{a(\bs k,s) + \alpha(\bs k,s,\bs r,t)},
\end{align}
where $\alpha$ is the radiated field.
Assuming $\bs p(t)= \bs p_0 e^{-i\omega_0 t} + \textrm{c.c.}$, where
$\textrm{c.c.}$ denotes the complex conjugate, $\alpha$ becomes
\begin{align}
\alpha(\bs k,s,\bs r,T) &=
\sqrt{\frac{\omega_0}{2(2\pi)^3\hbar\epsilon_0}}
e^{-i\bs k\cdot\bs r}\bs\varepsilon^*(\bs k,s)
\cdot
\nonumber\\&\quad
\left[\bs p_0 e^{i(\omega-\omega_0)T/2}\frac{\sin(\omega-\omega_0)T/2}{(\omega-\omega_0)/2}
\right.
\nonumber\\&\quad
\left.
+\bs p_0^*e^{i(\omega+\omega_0)T/2}\frac{\sin(\omega+\omega_0)T/2}{(\omega+\omega_0)/2}
\right],
\label{alpha}
\end{align}
which indicates that only the optical modes with $\omega(\bs k) =
\omega_0$ grow in time, corresponding to the far field, while all the
other near-field modes with $\omega(\bs k) \neq \omega_0$
oscillate. The $\omega(\bs k) = \omega_0$ relation specifies
the spatial frequencies available to the far optical fields
\cite{goodman,heintzmann09}.
Assuming $T\gg 2\pi/\omega_0$, such that
$\sin^2[(\omega\pm\omega_0)T/2]/[(\omega\pm\omega_0)/2]^2\approx 2\pi
T\delta(\omega\pm\omega_0)$, using the identity $\sum_s
\varepsilon_\mu(\bs k,s)\varepsilon_\nu^*(\bs k,s) = \delta_{\mu\nu} -
k_\mu k_\nu/|\bs k|^2$ \cite{mandel}, and switching to the spherical
coordinate system for $\bs k$, it can be shown that the average number
of radiated photons for an initial vacuum state is
\begin{align}
N &\equiv \sum_s \int d^3k \abs{\alpha(\bs k,s,\bs r,T)}^2
\approx \frac{|\bs p_0|^2\omega_0^3 T}{3\pi \hbar\epsilon_0c^3}.
\label{N}
\end{align}
The far-field limit ($\omega_0T \to\infty$) will be assumed hereafter.

I now focus on two representative cases: a linearly polarized dipole
with $\bs p_0 = p_0\hat{\bs z}$ and a circularly polarized dipole with
$\bs p_0 = p_0(\hat{\bs x}+i\hat{\bs y})/\sqrt{2}$.  Taking the
unknown parameters to be $\bs r$, the generators for $X = (x,y,z)$ can
be expressed, after some algebra, as
\begin{align}
\Delta g_\mu(\bs r) &= -\frac{\sqrt{2}}{W_\mu}\Delta P_\mu(\bs r), 
\quad \mu \in \{x,y,z\},
\label{dg}
\\
\Delta P_\mu(\bs r) &\equiv \frac{1}{\sqrt{2}i}\Bk{\Delta b_\mu(\bs
    r) -\Delta b_\mu^\dagger(\bs r)},
\end{align}
where $\Delta b_\mu$ is a normalized annihilation operator defined as
\begin{align}
\Delta b_\mu(\bs r) \equiv W_\mu\sum_s \int d^3k\,
\Bk{-ik_\mu \alpha^*(\bs k,s,\bs r,T)}
\Delta a(\bs k,s),
\end{align}
such that $[\Delta b_\mu(\bs r),\Delta b_\nu^\dagger(\bs r)]=
\delta_{\mu\nu}$, and the normalization constants $W_\mu$ are
\begin{align}
W_\mu &\equiv \Bk{\sum_s \int d^3k\, k_\mu^2 \abs{\alpha_\mu(\bs k,s,\bs r,T)}^2}^{-1/2}.
\label{Wmu}
\end{align}
The $d^3k$ integrals can again be computed with the help of the
far-field assumption and spherical coordinates. The results depend on
$\bs p_0$; for $\bs p_0 = p_0\hat{\bs z}$,
\begin{align}
W_x &= W_y = \sqrt{\frac{5}{2}}\frac{\lambda_0}{2\pi\sqrt{N}},
&
W_z &= \frac{\sqrt{5}\lambda_0}{2\pi\sqrt{N}},
\label{Wxyz1}
\end{align}
and for $\bs p_0 = p_0(\hat{\bs x}+i\hat{\bs y})/\sqrt{2}$,
\begin{align}
W_x &= W_y = \sqrt{\frac{10}{3}}\frac{\lambda_0}{2\pi\sqrt{N}},
&
W_z &= \sqrt{\frac{5}{2}}\frac{\lambda_0}{2\pi\sqrt{N}},
\label{Wxyz2}
\end{align}
but the important point here is that they are all on the order of
$\lambda_0/\sqrt{N}$, where $\lambda_0 \equiv 2\pi c/\omega_0$ is the
free-space wavelength.  The QFI becomes
\begin{align}
J_{\mu\nu}(\bs r) &= \frac{8}{W_\mu^2}\Avg{\Delta P_\mu(\bs r)\Delta P_\nu(\bs r)}. 
\end{align}
For an initial vacuum state (or any coherent state), $\avg{\Delta
  P_\mu(\bs r)\Delta P_\nu(\bs r)}=\delta_{\mu\nu}/2$, and the QCRB is
hence
\begin{align}
J_{\mu\nu}(\bs r) &= \frac{4}{W_\mu^2}\delta_{\mu\nu},
&
\Sigma_{\mu\mu}(\bs r) &\ge  \frac{W_\mu^2}{4},
\label{qsnl}
\end{align}
meaning that the quantum resolution limit in terms of the
root-mean-square error $\sqrt{\Sigma_{\mu\mu}}$ is on the order of
$\lambda_0/\sqrt{N}$. I call this limit the quantum shot-noise
limit. Generalization to lossless media is straightforward and results
simply in $\lambda_0$ being replaced by the wavelength in the
medium. Sec.~\ref{sec_single1} shows that a single-photon source also
obeys this limit with repeated trials.

Assuming uncorrelated photons,
Refs.~\cite{bobroff,thompson02,ober,centroid} derived a similar limit
in the form of $\sigma/\sqrt{N}$, where $\sigma$ is the width of the
imaging point-spread function. While their limit also scales as
$1/\sqrt{N}$, all those analyses assume the paraxial approximation and
measurement by a photon-counting camera, whereas the quantum
shot-noise limit here is valid for any numerical aperture and any
measurement, including common methods such as photon counting,
homodyne/heterodyne detection, and digital holography. The limit in
Refs.~\cite{bobroff,thompson02,ober,centroid} also implies that the
quantum shot-noise limit is reasonably tight, as the camera
measurement with suitable postprocessing can at least follow the
quantum-optimal shot-noise scaling. Sec.~\ref{sec_enhance} shows that
homodyne measurement with a special local-oscillator field can also
approach the quantum limit if the radiation is coherent.

For a concrete numerical example, consider the semiclassical paraxial
analysis of conventional single-molecule microscopy by Ober \textit{et
  al.}~\cite{ober}, who used the classical Cram\'er-Rao bound and a
shot-noise assumption to derive a limit of $2.301$~nm on the
root-mean-square localization error for a free-space wavelength of
$520~$nm, a numerical aperture of $1.4$, a photon collection
efficiency of 0.033, a photon flux of $2\times 10^6~$s$^{-1}$, an
acquisition time of $0.01$~s. If the efficiency were $1$, their limit
would become $0.418$~nm. In comparison, if I take the refractive index
of the immersion oil to be $1.52$, $\lambda_0 =
520$~nm$/1.52
= 342~$nm
to be the wavelength in the medium, and the photon number to be
$N = 2\times 10^4$, the quantum shot-noise limit according to
Eqs.~(\ref{Wxyz1}), (\ref{Wxyz2}), and (\ref{qsnl}) is
$\lambda_0/(2\pi\sqrt{N}) = 0.385$~nm times a constant factor close to
1.

It remains to be seen whether superoscillation techniques are
similarly efficient, but the key point here is that, since the quantum
bound is valid for any measurement and conventional methods can
already get close to it, no other measurement technique is able to
offer any significant advantage in resolution enhancement over the
conventional methods.

\section{Quantum enhancement}
\label{sec_enhance}
Even though the source is classical, quantum enhancement is possible
if the initial state $\ket{\psi}$ is nonclassical, as I now show.  If
$\Delta g_\mu$ were independent of the parameter, the accuracy could
be enhanced by squeezing and measuring the conjugate quadrature
\cite{braunstein96}. Although $\Delta g_\mu(\bs r)$ depends on the
unknown $\bs r$ here, the radiated field can be approximated as
$\alpha(\bs k,s,\bs r,T) \approx \alpha(\bs k,s,\bs r_0,T)$, resulting
in $\Delta g_\mu(\bs r) \approx \Delta g_\mu(\bs r_0)$, provided that
\begin{align}
  \abs{\bs r-\bs r_0} \ll \lambda_0
\end{align}
with respect to a known reference position $\bs r_0$.  The
acquisition of such prior information will require a fixed amount of
overhead resource, but once it is done, one can squeeze the quadrature
\begin{align}
\Delta Q_\mu(\bs r_0) &\equiv 
\frac{1}{\sqrt{2}}\Bk{\Delta b_\mu(\bs r_0)+\Delta b_\mu^\dagger(\bs r_0)}
\end{align}
in the initial state and perform a homodyne measurement of
$\Delta Q_\mu(\bs r_0)$ to estimate $\bs r$ much more
accurately. Since $[\Delta Q_\mu(\bs r_0), \Delta Q_\nu(\bs r_0)] =
0$, all three quadratures can be squeezed and measured simultaneously
in principle. The estimation error becomes
\begin{align}
  \Sigma_{\mu\mu}(\bs r) &\approx \frac{W_\mu^2}{2}\Avg{\Delta
    Q_\mu^2(\bs r_0)},
\end{align}
and the error reduction below the shot-noise limit is determined by
the squeezing factor, which is limited by the average photon number
$N_0$ in the initial state (not to be confused with $N$). Using
$\Avg{\Delta Q_\mu^2(\bs r_0)}+\Avg{\Delta P_\mu^2(\bs r_0)} \le
2N_0+1$ and the uncertainty relation $\avg{\Delta Q_\mu^2}\avg{\Delta
  P_\mu^2} \ge 1/4$, it can be shown that
\begin{align}
\Avg{\Delta Q_\mu^2(\bs r_0)} &\ge \frac{f(N_0)}{2},
\quad
\Avg{\Delta P_\mu^2(\bs r_0)} \le \frac{1}{2f(N_0)},
\\
f(N_0) &\equiv (2N_0+1)\Bk{1-\sqrt{1-(2N_0+1)^{-2}}},
\end{align}
where 
\begin{align}
f(0) &= 1,
&
f(N_0)&\approx \frac{1}{4N_0} \textrm{ for } N_0\gg 1.
\end{align}
With a zero-mean minimum-uncertainty state and all initial photons in
the $\Delta b_\mu(\bs r_0)$ mode, the estimation error becomes
\begin{align}
  \Sigma_{\mu\mu}(\bs r) &\approx \frac{W_\mu^2}{2}\Avg{\Delta Q_\mu^2(\bs r_0)}
  =\frac{W_\mu^2}{4}f(N_0).
\end{align}
The enhancement factor $f(N_0)$ is optimal, as the QCRB can be further
bounded by
\begin{align}
\Sigma_{\mu\mu}(\bs r) &\ge 
J_{\mu\mu}^{-1}(\bs r) =
\frac{W_\mu^2}{8\avg{\Delta P_\mu^2(\bs r)}} \ge \frac{W_\mu^2}{4}f(N_0).
\end{align}
This means that squeezed light with average photon number $N_0$ can
beat the quantum shot-noise limit to the mean-square error by roughly
a factor of $N_0$.

The optical mode to be squeezed has a profile
$ik_\mu\alpha(\bs k,s,\bs r_0,T)$. This means that, in real space, the
electric field profile of the mode should be the spatial derivative of
the radiated field. This kind of squeezing and measurement has
actually been demonstrated experimentally, albeit in the paraxial
regime, by Taylor \textit{et al.}\ in the context of particle tracking
\cite{taylor2013}, where the weak scatterer under a strong pump can be
modeled as a classical source, similar to the implementation of field
displacement by a beam splitter \cite{paris96}, and the spatial mode
profile of the squeezed light and the local oscillator is a spatial
derivative of the scattered field. To realize an enhancement in
practice, accurate phase locking of the squeezed light and the local
oscillator to the radiated field and a high measurement efficiency are
crucial. Phase locking cannot be achieved with incoherent point
sources such as fluorescent markers, but can be done with dielectric
particles via Rayleigh scattering \cite{taylor2013} or second-harmonic
nanoparticles \cite{pu08,hsieh09}; the latter are especially promising
for biological imaging applications \cite{dempsey}.

\begin{figure}[htbp]
\centerline{\includegraphics[width=0.4\textwidth]{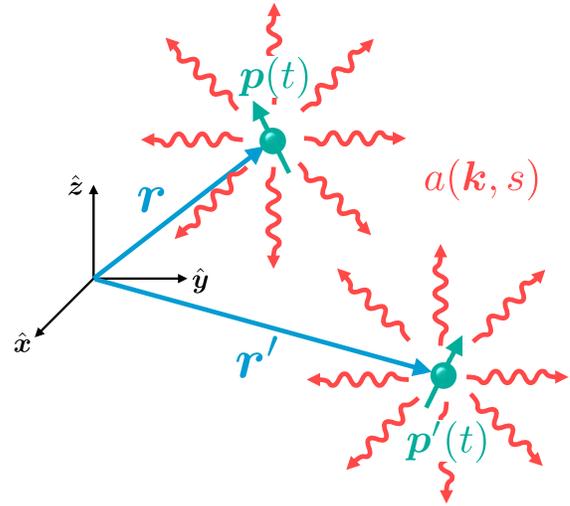}}
\caption{Two classical point sources with dipole
  moments $\bs p(t)$ and $\bs p'(t)$ at $\bs r$ and $\bs r'$ with
  quantum optical radiation.}
\label{2source}
\end{figure}

\section{Two classical point sources}
\label{sec_twosources}
\begin{figure}[htbp]
\centerline{\includegraphics[width=0.4\textwidth]{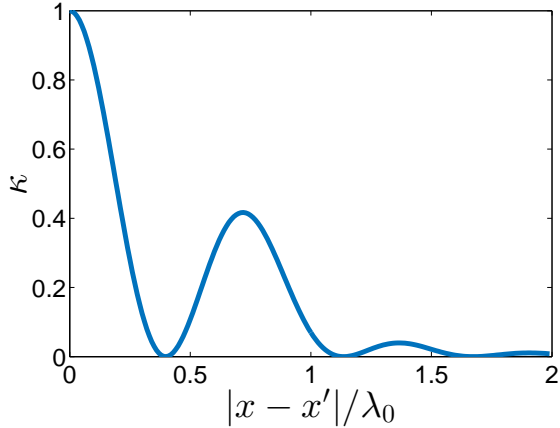}}
\caption{Plot of the resolution degradation factor $\kappa$ versus the
  true separation $|x-x'|/\lambda_0$ between two in-phase point
  sources, assuming
  $\bs p = \bs p' = \bs p_0 e^{-i\omega_0t}+\textrm{c.c.}$,
  $T\gg 2\pi/\omega_0$, $\bs p_0 = p_0\hat{\bs x}$, $y=y'$, and
  $z=z'$. At $|x-x'| = 0$, the Fisher information matrix is singular
  \cite{stoica,rotnitzky}. $\kappa$ remains the same for two
  out-of-phase sources with otherwise the same assumptions.}
\label{resolution_dz}
\end{figure}

Next, consider two classical point sources at $\bs r$ and $\bs r'$, as
shown in Fig.~\ref{2source}. The Hamiltonian is now
\begin{align}
H(\bs r,\bs r',t) &= H_F + H_I(\bs r,\bs r',t),
\\
H_I(\bs r,\bs r',t) &= 
-\bs p(t)\cdot\bs E(\bs r)-\bs p'(t)\cdot\bs E(\bs r').
\end{align}
The Heisenberg picture of $a(\bs k,s)$ becomes $a(\bs k,s,t) =
e^{-i\omega t}[a(\bs k,s) + \alpha(\bs k,s,\bs r,t) +\alpha'(\bs
k,s,\bs r',t)]$, where $\alpha$ and $\alpha'$ are the radiated fields
from the two sources, $\alpha$ is the same as before, and $\alpha'$
has the same expression as $\alpha$ except that $\bs p$ is replaced by
$\bs p'$ and $\bs r$ by $\bs r'$. One can then follow the preceding
procedure to obtain the QCRB for estimating $\bs r$ and $\bs r'$. To
highlight the important physics, however, consider here the estimation
of just two parameters $X = (x,x')$. The generators $\Delta g_x$ and
$\Delta g_{x'}$ may not commute,
and the QFI matrix for an initial vacuum or any coherent state now has
off-diagonal components:
\begin{align}
J_{xx'}(X) &= J_{x'x}(X)
\nonumber\\
&= 4\real \sum_s \int d^3k\, k_x^2 
\alpha^*(\bs k,s,\bs r,T)\alpha'(\bs k,s,\bs r',T),
\end{align}
while $J_{xx}$ remains the same and $J_{x'x'}$ has a similar
expression to $J_{xx}$. $J_{xx}$ and $J_{x'x'}$ still obey a
shot-noise scaling with the average photon number, but the nonzero
off-diagonal components mean that the parameters act as nuisance
parameters to each other, and the QCRB with respect to, say, $x$ is
always raised:
\begin{align}
\Sigma_{xx}(X) &\ge \frac{1}{J_{xx}[1-\kappa(X)]},
\end{align}
where the resolution degradation factor, defined as
\begin{align}
\kappa(X) &\equiv \frac{J_{xx'}^2(X)}{J_{xx}J_{x'x'}}
=
\frac{(\real \sum_s \int d^3k\, k_x^2 
\alpha^*\alpha')^2}
{\sum_s \int d^3k\, k_x^2 
|\alpha|^2\sum_s \int d^3k\, k_x^2 |\alpha'|^2},
\end{align}
is within the range $0\le \kappa\le 1$ and determined by the overlap
between the two differential mode profiles.  The nuisance-parameter
effect generalizes the Rayleigh criterion and other classical results
\cite{ram} by revealing a fundamental measurement-independent
degradation of resolution for two point sources with overlapping
radiation. For example, Fig.~\ref{resolution_dz} plots $\kappa$
against $|x-x'|/\lambda_0$, assuming
$\bs p = \bs p' = \bs p_0 e^{-i\omega_0t}+\textrm{c.c.}$,
$T\gg 2\pi/\omega_0$, $\bs p_0 = p_0\hat{\bs x}$, $y=y'$, and
$z=z'$. $\kappa\approx 0$ for $|x-x'| \gg \lambda_0$, as expected, but
it approaches $1$ and leads to a diverging QCRB when
$|x-x'|\ll \lambda_0$. Sec.~\ref{sec_partial2} shows that the
degradation effect should still exist for two partially coherent
sources.

The degradation effect can be avoided by minimizing the overlap before
each source is located independently. The overlap can be reduced by
making the radiated fields separate in space, time, frequency,
quadrature, or polarization; time multiplexing of point sources has
especially been the key driver in current superresolution microscopy
\cite{hell,betzig,moerner}.

Note that $\kappa$ also depends on the relative phase between $\alpha$
and $\alpha'$. For example, under the assumptions in the caption of
Fig.~\ref{resolution_dz}, it can be shown that the QFI matrix
transformed with respect to the average position $(x+x')/2$ and the
separation $x-x'$ is diagonal. The QFI component with respect to the
average position still obeys a shot-noise scaling, while the increase
in $\kappa$ can be traced to the increased error in estimating the
separation. Similarly, when $\alpha$ and $\alpha'$ are $180^\circ$ out
of phase but otherwise obey the same assumptions, $\kappa$ remains the
same but its increase is now due to the increased error for the
average position. An interesting scenario occurs when $\alpha$ and
$\alpha'$ are $90^\circ$ out of phase, the two fields are in
orthogonal quadratures, $\kappa$ becomes zero, and they can be
measured separately using homodyne or heterodyne detection. This
phenomenon suggests that structured illumination
\cite{gustafsson99,gustafsson,heintzmann,heintzmann09} can be used to excite the
sources, such that their relative phase and amplitude can be
controlled to some degree and the overlap can be reduced for certain
ranges of parameters.

 When the overlap is unavoidable or when the generators do
not commute, heterodyne measurements can still be used to measure both
quadratures of $a(\bs k,s,T)$ and should have a classical Fisher
information within a factor of $1/2$ of the QFI. Quantum enhancement
may also be possible using entangled squeezed states \cite{genoni};
the specific experimental design will be left for future studies.

\section{Bayesian quantum Cram\'er-Rao bound
with nuisance parameters}
\label{sec_qcrb_nuisance}
Incoherent sources are characterized by the statistical fluctuations
of the fields \cite{mandel}. For point-source radiation, incoherence
originates from the randomness of the amplitude, phase, and direction
of the point dipole.  In the context of statistical inference, these
random variables are most suitably modeled as nuisance parameters.
There are many ways to generalize the Cram\'er-Rao bounds when
nuisance parameters are present \cite{bell}.  The previous sections
show one way, which includes the nuisance parameters as part of the
wanted parameters $X$.  To derive tighter bounds for other types of
nuisance parameters, here I start with a Bayesian QCRB and generalize
a classical approach by Miller and Chang \cite{bell,miller_chang}. Let
$Z$ be a set of nuisance parameters, and suppose first that $Z$ is
given.  The Bayesian estimation error matrix is
\begin{align}
\bar\Sigma_{\mu\nu}(Z) &\equiv \int dX dY P(Y|X,Z)P_{X|Z}(X|Z)
\Bk{\tilde X_\mu(Y,Z)-X_\mu}
\nonumber\\&\quad\times
\Bk{\tilde X_\nu(Y,Z)-X_\nu},
\end{align}
where $P_{X|Z}(X|Z)$ is the prior distribution of $X$ conditioned on
$Z$. Note that this Bayesian definition of error regards $X$ as a
random parameter by averaging over its prior and is different from the
frequentist definition in Eq.~(\ref{Sigma}) \cite{vantrees}.  A
Bayesian quantum Cram\'er-Rao bound valid for any estimator is given
by \cite{yuen_lax,twc}
\begin{align}
\bar\Sigma(Z) &\ge \bar J^{-1}(Z),
\label{bayes_qcrb}
\\
\bar J(Z) &= \expect_{X|Z}\Bk{J(X|Z)} + j(Z),
\end{align}
where $J(X|Z)$ is the same QFI as before,
except that it is now conditioned on $Z$, $\expect_{X|Z}$ denotes
expectation over $P_{X|Z}(X|Z)$, and $j(Z)$ is a prior Fisher
information defined as
\begin{align}
j_{\mu\nu}(Z) &\equiv \int dX P_{X|Z}(X|Z)\Bk{\parti{}{X_\mu} \ln P_{X|Z}(X|Z)}
\nonumber\\&\quad\times
\Bk{\parti{}{X_\nu} \ln P_{X|Z}(X|Z)}.
\end{align}
If $Z$ is a random parameter with prior distribution given by
$P_Z(Z)$, the estimation error is
\begin{align}
\Pi_{\mu\nu} &\equiv \expect_Z\Bk{\bar{\Sigma}'(Z)},
\\
\bar{\Sigma}'(Z) &\equiv \int dX dY P(Y|X,Z)P_{X|Z}(X|Z)
\Bk{\tilde X_\mu(Y)-X_\mu}
\nonumber\\&\quad\times
\Bk{\tilde X_\nu(Y)-X_\nu},
\end{align}
where $\expect_Z$ denotes expectation over $P_Z$ and the estimator
$\tilde X_\mu(Y)$ can no longer depend on $Z$.  The lower bound in
Eq.~(\ref{bayes_qcrb}) still holds for $\bar{\Sigma}'(Z)$,
so one can obtain a lower bound on $\Pi$ given by
\begin{align}
\Pi &\ge \expect_Z\Bk{\bar J^{-1}(Z)}.
\label{newbound}
\end{align}
The important mathematical feature of this bound is that the
expectation with respect to the nuisance parameter $Z$ is taken after
the inverse of the conditional QFI matrix. This can sometimes lead to
a tighter bound than a QCRB that includes $Z$ as part of $X$. Note
also that this Bayesian bound is valid for any estimator, not just the
unbiased ones, unlike the claim in Ref.~\cite{miller_chang}. The
tightness of the bound should depend on whether the nuisance
parameters can be accurately estimated from the measurements.

\section{One partially coherent source}
\label{sec_partial1}
I now use the new bound to study partially coherent sources.  First,
consider the example of one point source in Sec.~\ref{sec_onesource},
but suppose that the complex dipole amplitude $p_0$ is
unknown. Assuming $Z = p_0$, the quantum state before measurement is
\begin{align}
\rho &= \int d^2p_0 P_Z(p_0) U(X,p_0,T)\rho_0U^\dagger(X,p_0,T).
\end{align}
If $\rho_0$ is a vacuum state, $U\rho_0U^\dagger$ is a coherent state,
and $\rho$ is a classical mixed state of light with $P_Z(p_0)$
determining the Sudarshan-Glauber P function \cite{mandel}. The
random $p_0$ therefore gives rise to a classical partially coherent
source model. For an initial vacuum, $\bar J(p_0)$ is given by
\begin{align}
  \bar J_{\mu\nu}(p_0) &= \frac{4}{W_\mu^2(p_0)}\delta_{\mu\nu} +j_{\mu\nu} 
= \frac{N(p_0)}{C_\mu\lambda_0^2}\delta_{\mu\nu} + j_{\mu\nu},
\end{align}
where $W_\mu$ and $N$ now depend on the unknown dipole moment and
$C_\mu$ is a constant on the order of 1 that can be determined from
Eqs.~(\ref{Wxyz1}) or (\ref{Wxyz2}).  Assuming that $j$ is diagonal
and independent of $p_0$ and taking the inverse and then the
expectation according to Eq.~(\ref{newbound}), I obtain
\begin{align}
\Pi_{\mu\mu} &\ge \expect_{p_0}\Bk{\frac{1}{4/W_\mu^2(p_0) + j_{\mu\mu}}}
=
\expect_{p_0}
\Bk{\frac{1}{ N(p_0)/(C_\mu\lambda_0^2) + j_{\mu\mu}}}.
\end{align}
For example, if $P_Z(p_0)$ corresponds to the P function of a
thermal source, the bound can be written in terms of the average
radiated photon number $\bar N$ as
\begin{align}
&\quad \expect_{p_0}
\Bk{\frac{1}{N(p_0)/(C_\mu \lambda_0^2) + j_{\mu\mu}}}
\nonumber\\
&= \frac{C_\mu \lambda_0^2}{\bar N} \int_0^\infty dN 
\exp\bk{-\frac{N}{\bar N}}\frac{1}{N+C_\mu\lambda_0^2 j_{\mu\mu}}
\\
&\approx \frac{C_\mu\lambda_0^2}{\bar N}  
\ln \frac{\bar N}{C_\mu \lambda_0^2 j_{\mu\mu}},
\quad
\frac{\bar N}{C_\mu\lambda_0^2} \gg j_{\mu\mu}.
\end{align}
An alternative Bayesian QCRB can be obtained by including $p_0$ as
part of $X$. In that case, the off-diagonal QFI elements between $p_0$
and $\bs r$ turn out to be zero, and the expectation with respect to
$p_0$ is taken before the inverse, leading to a bound given by
$C_\mu\lambda_0^2/\bar N$. The separate treatment of $p_0$ as nuisance
parameter here involves taking the expectation after the inverse,
giving rise to an additional factor $\sim \ln\bar N$ and thus a
tighter bound for large $\bar N$.  In general, Jensen's inequality can
be used to show that the shot-noise scaling with respect to the
average photon number $\expect_{p_0}[N(p_0)]$ cannot be beat for any
nonnegative P function.

\section{Two partially coherent sources}
\label{sec_partial2}
Next, consider the example of two point sources in
Sec.~\ref{sec_twosources}, and $Z = (\bs p_0,\bs p_0')$ is now assumed
to be unknown to model partially coherent sources. Assuming again that
$j$ is diagonal and $X = (x,x')$ is independent of $Z$,
\begin{align}
\bar J_{xx}(\bs p_0) &= \frac{4}{W_x^2(\bs p_0)} + j_{xx},
\\
\bar J_{x'x'}(\bs p_0') &= \frac{4}{W_{x'}^2(\bs p_0')} + j_{x'x'},
\\
\bar J_{xx'}(\bs p_0,\bs p_0') &= \expect_X\Bk{J_{xx'}(\bs p_0,\bs p_0')}.
\end{align}
$\bar J_{xx'}$ is now the expectation of $J_{xx'}$ over $X = (x,x')$,
conditioned on the dipole moments. If the two point sources are
\textit{a priori} known to be close relative to $\lambda_0$, $\bar
J_{xx'}(\bs p_0,\bs p_0')$ can still have a significant magnitude for
certain $(\bs p_0,\bs p_0')$. The bound given by Eq.~(\ref{newbound})
becomes
\begin{align}
\Pi_{xx} &\ge \expect_{(\bs p_0,\bs p_0')}
\BK{\frac{1}{\bar J_{xx}(\bs p_0)[1-\bar\kappa(\bs p_0,\bs p_0')]}},
\label{Pi_xx}
\end{align}
where a new resolution degradation factor is defined as
\begin{align}
  \bar\kappa(\bs p_0,\bs p_0') &\equiv \frac{\bar J_{xx'}^2(\bs
    p_0,\bs p_0')}{\bar J_{xx}(\bs p_0)\bar J_{x'x'}(\bs p_0')}.
\end{align}
The important point here is that $\bar\kappa(\bs p_0,\bs p_0')$ can
still be close to $1$ for certain values of $(\bs p_0,\bs p_0')$ if
the two sources are \textit{a priori} known to be close to each other
relative to $\lambda_0$, $1/[1-\bar\kappa(\bs p_0,\bs p_0')]\gg 1$ is
possible, and the expectation over $(\bs p_0,\bs p_0')$ will then be
dominated by those large values. In other words, the resolution
degradation effect derived for coherent sources must still exist for
partially coherent sources if their radiated fields may have
significant overlap.

An alternative Bayesian QCRB that includes $(\bs p_0,\bs p_0')$ as
part of $X$ can again be computed, but at some stage it involves
taking the expectation of $\bar J_{xx'}$ with respect to
$(\bs p_0,\bs p_0')$ before the inverse is taken. For incoherent
sources, this can reduce the off-diagonal components significantly;
the resulting bound, while still valid, would be less tight and no
longer demonstrate the resolution degradation effect.

\section{Single-photon source}
\label{sec_single1}
Consider now an initially excited two-level atom in free space. A
detailed analysis of atom-photon interaction is formidable
\cite{mandel,atom-photon}, but when the initial optical state is
vacuum, spontaneous emission can be treated more easily, as the atom
must decay to ground state in the long-time limit and the final
optical state must contain exactly one photon. Using the continuous
Fock space \cite{mandel}, the final optical state in the Schr\"odinger
picture can be written with respect to the vacuum $\ket{0}$ as
\begin{align}
\ket{\Psi} &= c^\dagger\ket{0},
&
c^\dagger &\equiv \sum_s \int d^3k \phi(\bs k,s)a^\dagger(\bs k,s),
\end{align}
where 
\begin{align}
\phi(\bs k,s) &= \Avg{\bs k,s|\Psi} =
\bra{0}a(\bs k,s)\ket{\Psi}
\end{align}
is the one-photon configuration-space amplitude.  Following
Chap.~III.C of Ref.~\cite{atom-photon}, it can be expressed as
\begin{align}
\phi(\bs k,s) &= \frac{\bra{\bs k,s}\otimes \bra{g}H_I\ket{e}\otimes\ket{0}}
{\hbar[(\omega-\tilde{\omega}_0)+i/(2T_1)]}e^{-i\omega T},
\\
H_I &= i\omega_0\bk{\bs\mu_{12}\sigma-\bs\mu_{12}^*\sigma^\dagger}
\cdot \bs A(\bs r),
\\
\bs A(\bs r) &= 
\sum_s \int d^3k\sqrt{\frac{\hbar}{2(2\pi)^3\omega\epsilon_0}}
\Bk{a(\bs k,s)\bs\varepsilon(\bs k,s) e^{i\bs k\cdot \bs r}+\textrm{h.c.}},
\\
T_1 &= \frac{3\pi\hbar\epsilon_0 c^3}{|\bs\mu_{12}|^2\omega_0^3},
\end{align}
where $\ket{e}$ and $\ket{g}$ are the excited and ground atomic
states, $\omega_0$ is the atomic resonance frequency, $\tilde\omega_0$
is the Lamb-shifted atomic frequency, $T_1$ is the decay time,
$\bs\mu_{12}$ is the off-diagonal atomic dipole moment, and
$\sigma\equiv \ket{g}\bra{e}$ is the atomic lowering operator. The
result is
\begin{align}
\phi(\bs k,s) &= 
\frac{1}{i(\omega-\tilde\omega_0)+1/(2T_1)}
\sqrt{\frac{\omega_0^2}{2(2\pi)^3\hbar\omega\epsilon_0}}
\bs\mu_{12}\cdot\bs\varepsilon^*(\bs k,s) 
\nonumber\\&\quad\times e^{-i\bs k\cdot\bs r-i\omega T}.
\end{align}
Consider the fidelity
\begin{align}
F &= \abs{\bra{0}c(X)c^\dagger(X+\delta X)\ket{0}}^2
\\
&= \abs{\Bk{c(X),c^\dagger(X+\delta X)}}^2
\\
&\approx  1+\sum_{\mu,\nu}\delta X_\mu\delta X_\nu
\left\{\real \Bk{c,\parti{^2c^\dagger}{X_\mu\partial X_\nu}}
\right.
\nonumber\\&\quad
\left.+\imag \Bk{c,\parti{c^\dagger}{X_\mu}}\imag \Bk{c,\parti{c^\dagger}{X_\nu}}\right\},
\end{align}
where the fact 
\begin{align}
\real \bra{0}c \parti{c^\dagger}{X_\mu}\ket{0}
&= \real \Bk{c,\parti{c^\dagger}{X_\mu}} = 0
\end{align}
due to $\bra{0}c(X+\delta X)c^\dagger(X+\delta X)\ket{0} =
\bra{0}c(X)c^\dagger(X)\ket{0} = 1$ has been used.  It can further be 
shown that
\begin{align}
\Bk{c,\parti{c^\dagger}{X_\mu}} &=
\sum_s \int d^3k \phi^*(\bs k,s)\parti{\phi(\bs k,s)}{X_\mu} = 0,
\end{align}
leading to a QFI given by
\begin{align}
J_{\mu\nu} &= -4\real
\Bk{c,\parti{^2c^\dagger}{X_\mu\partial X_\nu}}
\\
&= -4\real \sum_s \int d^3k 
\phi^*(\bs k,s)\parti{^2\phi(\bs k,s)}{X_\mu\partial X_\nu}.
\\
&= 4\real\sum_s \int d^3k 
\frac{k_\mu k_\nu}{(\omega-\tilde\omega_0)^2+1/(4T_1^2)}
\frac{\omega_0^2}{2(2\pi)^3\hbar\omega\epsilon_0}
\nonumber\\&\quad\times
\abs{\bs\mu_{12}\cdot\bs\varepsilon^*(\bs k,s)}^2.
\end{align}
Assuming that the decay time is much longer than the optical period
($T_1 \gg 2\pi/\tilde\omega_0$) and the Lamb shift is much smaller
than the optical $\omega_0$ ($\tilde\omega_0 \approx \omega_0$), the
QFI becomes
\begin{align}
J_{\mu\nu}
&\approx 4\real\sum_s \int d^3k k_\mu k_\nu 2\pi T_1\delta(\omega-\omega_0)
\frac{\omega_0^2}{2(2\pi)^3\hbar\omega\epsilon_0}
\nonumber\\&\quad\times
\abs{\bs\mu_{12}\cdot\bs\varepsilon^*(\bs k,s)}^2.
\end{align}
This turns out to be identical to the QFI derived in
Sec.~\ref{sec_onesource} for an $N=1$ classical source:
\begin{align}
J_{\mu\nu} &=\frac{4\delta_{\mu\nu}}{N W_\mu^2} \sim 
\frac{\delta_{\mu\nu}}{\lambda_0^2},
&
\Sigma_{\mu\mu} &\ge J_{\mu\mu}^{-1}  = \frac{NW_\mu^2}{4} \sim \lambda_0^2,
\end{align}
where $N$ is defined in Eq.~(\ref{N}) and $W_\mu$ is defined in
Eq.~(\ref{Wmu}) and given by Eqs.~(\ref{Wxyz1}) or Eqs.~(\ref{Wxyz2}),
such that $NW_\mu^2$ is on the order of $\lambda_0^2$.  This result
shows that a single-photon source offers no fundamental advantage over
a classical source that emits one photon on average. Superresolution
beyond the classical Abbe-Rayleigh limit can still be obtained,
however, if the experiment can be repeated. The QFI is then multiplied
by the number of trials $M$, which is also the total number of emitted
photons, and the resulting QCRB is identical to that for a classical
source with $M$ replacing $N$.  The experiments reported by
Refs.~\cite{schwartz13,cui13,monticone} certainly involved a large
number of measurements of many photons in total, which can explain the
apparent superresolution, but it remains to be seen whether their
methods are accurate or efficient in estimating object parameters.

For two atoms with large separation ($|\bs r-\bs r'|\gg \lambda_0$),
the one-atom analysis is expected to be applicable to each atom
independently.  The analysis of two close atoms is much more
challenging because of atomic cooperative effects such as the Dicke
superradiance \cite{mandel} and the F\"orster resonance energy
transfer \cite{moerner}. Beyond the current assumption of spontaneous
emission, it will also be interesting, though highly nontrivial, to
analyze the interaction between two-level atoms and other states of
light, such as coherent states or squeezed states, and investigate
their quantum localization limits and the possibility of quantum
enhancement. 

\section{Conclusion}
I have derived quantum limits to point-source localization using
quantum estimation theory and the quantum theory of electromagnetic
fields.  These results not only provide general no-go theorems
concerning the microscope resolution, they should also motivate
further progress in microscopy through classical or quantum techniques
beyond the current assumptions.  For example, the presented theory may
be applied to other more exotic quantum states of light interacting
with quantum sources, such as multilevel atoms \cite{mandel,scully},
quantum dots \cite{michalet,schwartz13}, diamond defects
\cite{hall_diamond,cui13,monticone}, and second-harmonic nanoparticles
\cite{pu08,hsieh09,dempsey}. It is also possible to generalize the
current formalism for open quantum systems
\cite{escher,demkowicz,tsang_open,knysh14} to account for mixed
states, decoherence, optical losses, background noises, and imperfect
measurement efficiency. The phenomenon of fluorescence intermittency
or blinking represents another interesting challenge to the
statistical analysis, as a fluorescent source can turn itself on and
off randomly and the localization of two blinking sources offers
another route to superresolution \cite{lidke}.  If done with care, the
application of quantum information science to microscopy is destined
to yield sound insights and opportunities in both fields.

\section*{Funding information}
Singapore National Research Foundation (NRF-NRFF2011-07).

 \bibliography{point-source_localization_optica}

\begin{thebibliography}{10}
\newcommand{\enquote}[1]{``#1''}

\bibitem{born_wolf}
M.~Born and E.~Wolf, \emph{Principles of Optics: Electromagnetic Theory of
  Propagation, Interference and Diffraction of Light} (Cambridge University
  Press, Cambridge, 1999).

\bibitem{hell}
S.~W. Hell, \enquote{Far-field optical nanoscopy,} Science \textbf{316},
  1153--1158 (2007).

\bibitem{betzig}
E.~Betzig, G.~H. Patterson, R.~Sougrat, O.~W. Lindwasser, S.~Olenych, J.~S.
  Bonifacino, M.~W. Davidson, J.~Lippincott-Schwartz, and H.~F. Hess,
  \enquote{Imaging intracellular fluorescent proteins at nanometer resolution,}
  Science \textbf{313}, 1642--1645 (2006).

\bibitem{moerner}
W.~E. Moerner, \enquote{New directions in single-molecule imaging and
  analysis,} Proceedings of the National Academy of Sciences \textbf{104},
  12596--12602 (2007).

\bibitem{bobroff}
N.~Bobroff, \enquote{Position measurement with a resolution and noise-limited
  instrument,} Review of Scientific Instruments \textbf{57}, 1152--1157 (1986).

\bibitem{thompson02}
R.~E. Thompson, D.~R. Larson, and W.~W. Webb, \enquote{Precise nanometer
  localization analysis for individual fluorescent probes,} Biophysical Journal
  \textbf{82}, 2775--2783 (2002).

\bibitem{ober}
R.~J. Ober, S.~Ram, and E.~S. Ward, \enquote{Localization accuracy in
  single-molecule microscopy,} Biophysical Journal \textbf{86}, 1185--1200
  (2004).

\bibitem{ram}
S.~Ram, E.~S. Ward, and R.~J. Ober, \enquote{Beyond rayleigh's criterion: A
  resolution measure with application to single-molecule microscopy,}
  Proceedings of the National Academy of Sciences of the United States of
  America \textbf{103}, 4457--4462 (2006).

\bibitem{mandel}
L.~Mandel and E.~Wolf, \emph{Optical Coherence and Quantum Optics} (Cambridge
  University Press, Cambridge, 1995).

\bibitem{sheppard07}
C.~J.~R. Sheppard, \enquote{The optics of microscopy,} Journal of Optics A:
  Pure and Applied Optics \textbf{9}, S1 (2007).

\bibitem{berry_popescu}
M.~V. Berry and S.~Popescu, \enquote{Evolution of quantum superoscillations and
  optical superresolution without evanescent waves,} Journal of Physics A:
  Mathematical and General \textbf{39}, 6965 (2006).

\bibitem{zheludev08}
N.~I. Zheludev, \enquote{What diffraction limit?} Nature Materials \textbf{7},
  420--422 (2008).

\bibitem{huang_zheludev}
F.~M. Huang and N.~I. Zheludev, \enquote{Super-resolution without evanescent
  waves,} Nano Letters \textbf{9}, 1249--1254 (2009).

\bibitem{hyvarinen12}
H.~J. Hyv\"{a}rinen, S.~Rehman, J.~Tervo, J.~Turunen, and C.~J.~R. Sheppard,
  \enquote{Limitations of superoscillation filters in microscopy applications,}
  Opt. Lett. \textbf{37}, 903--905 (2012).

\bibitem{helstrom}
C.~W. Helstrom, \emph{Quantum Detection and Estimation Theory} (Academic Press,
  New York, 1976).

\bibitem{holevo}
A.~S. Holevo, \emph{Statistical Structure of Quantum Theory} (Springer-Verlag,
  Berlin, 2001).

\bibitem{helstrom70}
C.~W. Helstrom, \enquote{Estimation of object parameters by a quantum-limited
  optical system,} J. Opt. Soc. Am. \textbf{60}, 233--239 (1970).

\bibitem{kolobov}
M.~I. Kolobov, \enquote{The spatial behavior of nonclassical light,} Rev. Mod.
  Phys. \textbf{71}, 1539--1589 (1999).

\bibitem{fabre}
C.~{Fabre}, J.~B. {Fouet}, and A.~{Ma{\^i}tre}, \enquote{{Quantum limits in the
  measurement of very small displacements in optical images},} Optics Letters
  \textbf{25}, 76--78 (2000).

\bibitem{treps}
N.~Treps, N.~Grosse, W.~P. Bowen, C.~Fabre, H.-A. Bachor, and P.~K. Lam,
  \enquote{A quantum laser pointer,} Science \textbf{301}, 940--943 (2003).

\bibitem{barnett}
S.~Barnett, C.~Fabre, and A.~Maıtre, \enquote{Ultimate quantum limits for
  resolution of beam displacements,} The European Physical Journal D - Atomic,
  Molecular, Optical and Plasma Physics \textbf{22}, 513--519 (2003).

\bibitem{boto}
A.~N. Boto, P.~Kok, D.~S. Abrams, S.~L. Braunstein, C.~P. Williams, and J.~P.
  Dowling, \enquote{Quantum interferometric optical lithography: Exploiting
  entanglement to beat the diffraction limit,} Phys. Rev. Lett. \textbf{85},
  2733--2736 (2000).

\bibitem{li08}
H.~Li, V.~A. Sautenkov, M.~M. Kash, A.~V. Sokolov, G.~R. Welch, Y.~V.
  Rostovtsev, M.~S. Zubairy, and M.~O. Scully, \enquote{Optical imaging beyond
  the diffraction limit via dark states,} Phys. Rev. A \textbf{78}, 013803
  (2008).

\bibitem{boyd2012}
R.~W. Boyd and J.~P. Dowling, \enquote{Quantum lithography: status of the
  field,} Quantum Information Processing \textbf{11}, 891--901 (2012).

\bibitem{centroid}
M.~Tsang, \enquote{Quantum imaging beyond the diffraction limit by optical
  centroid measurements,} Phys. Rev. Lett. \textbf{102}, 253601 (2009).

\bibitem{shin}
H.~Shin, K.~W.~C. Chan, H.~J. Chang, and R.~W. Boyd, \enquote{Quantum spatial
  superresolution by optical centroid measurements,} Phys. Rev. Lett.
  \textbf{107}, 083603 (2011).

\bibitem{rozema}
L.~A. Rozema, J.~D. Bateman, D.~H. Mahler, R.~Okamoto, A.~Feizpour, A.~Hayat,
  and A.~M. Steinberg, \enquote{Scalable spatial superresolution using
  entangled photons,} Phys. Rev. Lett. \textbf{112}, 223602 (2014).

\bibitem{glm_imaging}
V.~Giovannetti, S.~Lloyd, L.~Maccone, and J.~H. Shapiro,
  \enquote{Sub-rayleigh-diffraction-bound quantum imaging,} Phys. Rev. A
  \textbf{79}, 013827 (2009).

\bibitem{nair_yen}
R.~Nair and B.~J. Yen, \enquote{Optimal quantum states for image sensing in
  loss,} Phys. Rev. Lett. \textbf{107}, 193602 (2011).

\bibitem{pirandola}
S.~Pirandola, \enquote{Quantum reading of a classical digital memory,} Phys.
  Rev. Lett. \textbf{106}, 090504 (2011).

\bibitem{perez12}
C.~A. P\'erez-Delgado, M.~E. Pearce, and P.~Kok, \enquote{Fundamental limits of
  classical and quantum imaging,} Phys. Rev. Lett. \textbf{109}, 123601 (2012).

\bibitem{hemmer12}
P.~R. Hemmer and T.~Zapata, \enquote{The universal scaling laws that determine
  the achievable resolution in different schemes for super-resolution imaging,}
  Journal of Optics \textbf{14}, 083002 (2012).

\bibitem{humphreys}
P.~C. Humphreys, M.~Barbieri, A.~Datta, and I.~A. Walmsley, \enquote{Quantum
  enhanced multiple phase estimation,} Phys. Rev. Lett. \textbf{111}, 070403
  (2013).

\bibitem{schwartz_oron}
O.~Schwartz and D.~Oron, \enquote{Improved resolution in fluorescence
  microscopy using quantum correlations,} Phys. Rev. A \textbf{85}, 033812
  (2012).

\bibitem{schwartz13}
O.~Schwartz, J.~M. Levitt, R.~Tenne, S.~Itzhakov, Z.~Deutsch, and D.~Oron,
  \enquote{Superresolution microscopy with quantum emitters,} Nano Letters
  \textbf{13}, 5832--5836 (2013).

\bibitem{cui13}
J.-M. Cui, F.-W. Sun, X.-D. Chen, Z.-J. Gong, and G.-C. Guo, \enquote{Quantum
  statistical imaging of particles without restriction of the diffraction
  limit,} Phys. Rev. Lett. \textbf{110}, 153901 (2013).

\bibitem{monticone}
D.~Gatto~Monticone, K.~Katamadze, P.~Traina, E.~Moreva, J.~Forneris,
  I.~Ruo-Berchera, P.~Olivero, I.~P. Degiovanni, G.~Brida, and M.~Genovese,
  \enquote{Beating the abbe diffraction limit in confocal microscopy via
  nonclassical photon statistics,} Phys. Rev. Lett. \textbf{113}, 143602
  (2014).

\bibitem{taylor2013}
M.~A. Taylor, J.~Janousek, V.~Daria, J.~Knittel, B.~Hage, H.-A. Bachor, and
  W.~P. Bowen, \enquote{Biological measurement beyond the quantum limit,}
  Nature Photonics \textbf{7}, 229--233 (2013).

\bibitem{wiseman_milburn}
H.~M. Wiseman and G.~J. Milburn, \emph{Quantum Measurement and Control}
  (Cambridge University Press, Cambridge, 2010).

\bibitem{vantrees}
H.~L. Van~Trees, \emph{Detection, Estimation, and Modulation Theory, Part I.}
  (John Wiley \& Sons, New York, 2001).

\bibitem{braunstein}
S.~L. Braunstein and C.~M. Caves, \enquote{Statistical distance and the
  geometry of quantum states,} Phys. Rev. Lett. \textbf{72}, 3439--3443 (1994).

\bibitem{tsang_nair}
M.~Tsang and R.~Nair, \enquote{Fundamental quantum limits to waveform
  detection,} Phys. Rev. A \textbf{86}, 042115 (2012).

\bibitem{paris}
M.~G.~A. Paris, \enquote{Quantum estimation for quantum technology,}
  International Journal of Quantum Information \textbf{7}, 125--137 (2009).

\bibitem{pasquale}
A.~De~Pasquale, D.~Rossini, P.~Facchi, and V.~Giovannetti, \enquote{Quantum
  parameter estimation affected by unitary disturbance,} Phys. Rev. A
  \textbf{88}, 052117 (2013).

\bibitem{fujiwara2006}
A.~Fujiwara, \enquote{Strong consistency and asymptotic efficiency for adaptive
  quantum estimation problems,} Journal of Physics A: Mathematical and General
  \textbf{39}, 12489 (2006).

\bibitem{yuen_lax}
H.~P. Yuen and M.~Lax, \enquote{Multiple-parameter quantum estimation and
  measurement of nonselfadjoint observables,} IEEE Transactions on Information
  Theory \textbf{19}, 740--750 (1973).

\bibitem{twc}
M.~Tsang, H.~M. Wiseman, and C.~M. Caves, \enquote{Fundamental quantum limit to
  waveform estimation,} Phys. Rev. Lett. \textbf{106}, 090401 (2011).

\bibitem{qzzb}
M.~Tsang, \enquote{{Z}iv-{Z}akai error bounds for quantum parameter
  estimation,} Phys. Rev. Lett. \textbf{108}, 230401 (2012).

\bibitem{qbzzb}
D.~W. {Berry}, M.~{Tsang}, M.~J.~W. {Hall}, and H.~M. {Wiseman}, \enquote{{The
  quantum Bell-Ziv-Zakai bounds and Heisenberg limits for waveform
  estimation},} ArXiv e-prints  (2014).

\bibitem{goodman}
J.~W. Goodman, \emph{Introduction to Fourier Optics} (Mc{G}raw-{H}ill, New
  York, 2004).

\bibitem{heintzmann09}
R.~Heintzmann and M.~G.~L. Gustafsson, \enquote{Subdiffraction resolution in
  continuous samples,} Nature Photonics \textbf{3}, 362--364 (2009).

\bibitem{braunstein96}
S.~L. Braunstein, C.~M. Caves, and G.~J. Milburn, \enquote{Generalized
  uncertainty relations: Theory, examples, and lorentz invariance,} Annals of
  Physics \textbf{247}, 135 -- 173 (1996).

\bibitem{paris96}
M.~G. Paris, \enquote{Displacement operator by beam splitter,} Physics Letters
  A \textbf{217}, 78 -- 80 (1996).

\bibitem{pu08}
Y.~Pu, M.~Centurion, and D.~Psaltis, \enquote{Harmonic holography: a new
  holographic principle,} Appl. Opt. \textbf{47}, A103--A110 (2008).

\bibitem{hsieh09}
C.-L. Hsieh, R.~Grange, Y.~Pu, and D.~Psaltis, \enquote{Three-dimensional
  harmonic holographic microcopy using nanoparticles as probes for cell
  imaging,} Opt. Express \textbf{17}, 2880--2891 (2009).

\bibitem{dempsey}
W.~P. Dempsey, S.~E. Fraser, and P.~Pantazis, \enquote{Shg nanoprobes:
  Advancing harmonic imaging in biology,} BioEssays \textbf{34}, 351--360
  (2012).

\bibitem{gustafsson99}
M.~G.~L. Gustafsson, D.~A. Agard, and J.~W. Sedat, \enquote{I5m: 3d widefield
  light microscopy with better than 100 nm axial resolution,} Journal of
  Microscopy \textbf{195}, 10--16 (1999).

\bibitem{gustafsson}
M.~G.~L. Gustafsson, \enquote{Surpassing the lateral resolution limit by a
  factor of two using structured illumination microscopy,} Journal of
  Microscopy \textbf{198}, 82--87 (2000).

\bibitem{heintzmann}
R.~Heintzmann and C.~G. Cremer, \enquote{Laterally modulated excitation
  microscopy: improvement of resolution by using a diffraction grating,} Proc.
  SPIE \textbf{3568}, 185--196 (1999).

\bibitem{genoni}
M.~G. Genoni, M.~G.~A. Paris, G.~Adesso, H.~Nha, P.~L. Knight, and M.~S. Kim,
  \enquote{Optimal estimation of joint parameters in phase space,} Phys. Rev. A
  \textbf{87}, 012107 (2013).

\bibitem{stoica}
P.~Stoica and T.~L. Marzetta, \enquote{Parameter estimation problems with
  singular information matrices,} Signal Processing, IEEE Transactions on
  \textbf{49}, 87--90 (2001).

\bibitem{rotnitzky}
A.~Rotnitzky, D.~R. Cox, M.~Bottai, and J.~Robins, \enquote{Likelihood-based
  inference with singular information matrix,} Bernoulli \textbf{6}, 243--284
  (2000).

\bibitem{bell}
H.~L. Van~Trees and K.~L. Bell, eds., \emph{Bayesian Bounds for Parameter
  Estimation and Nonlinear Filtering/Tracking} (Wiley-IEEE, Piscataway, 2007).

\bibitem{miller_chang}
R.~W. Miller and C.~B. Chang, \enquote{A modified cram{\'e}r-rao bound and its
  applications (corresp.),} IEEE Transactions on Information Theory
  \textbf{24}, 398--400 (1978).

\bibitem{atom-photon}
C.~Cohen-Tannoudji, J.~Dupont-Roc, and G.~Grynberg, \emph{Atom-Photon
  Interactions} (Wiley-VCH Verlag GmbH, Weinheim, 2004).

\bibitem{scully}
M.~O. Scully and M.~S. Zubairy, \emph{Quantum Optics} (Cambridge University
  Press, Cambridge, 1997).

\bibitem{michalet}
X.~Michalet, F.~F. Pinaud, L.~A. Bentolila, J.~M. Tsay, S.~Doose, J.~J. Li,
  G.~Sundaresan, A.~M. Wu, S.~S. Gambhir, and S.~Weiss, \enquote{Quantum dots
  for live cells, in vivo imaging, and diagnostics,} Science \textbf{307},
  538--544 (2005).

\bibitem{hall_diamond}
L.~T. Hall, G.~C.~G. Beart, E.~A. Thomas, D.~A. Simpson, L.~P. McGuinness,
  J.~H. Cole, J.~H. Manton, R.~E. Scholten, F.~Jelezko, J.~Wrachtrup,
  S.~Petrou, and L.~C.~L. Hollenberg, \enquote{High spatial and temporal
  resolution wide-field imaging of neuron activity using quantum nv-diamond,}
  Sci. Rep. \textbf{2}, 401 (2012).

\bibitem{escher}
B.~M. Escher, R.~L. de~Matos~Filho, and L.~Davidovich, \enquote{General
  framework for estimating the ultimate precision limit in noisy
  quantum-enhanced metrology,} Nature Physics \textbf{7}, 406--411 (2011).

\bibitem{demkowicz}
R.~{Demkowicz-Dobrza{\'n}ski}, J.~{Ko{\l}ody{\'n}ski}, and M.~{Gu{\c t}{\u a}},
  \enquote{{The elusive Heisenberg limit in quantum-enhanced metrology},}
  Nature Communications \textbf{3}, 1063 (2012).

\bibitem{tsang_open}
M.~Tsang, \enquote{Quantum metrology with open dynamical systems,} New Journal
  of Physics \textbf{15}, 073005 (2013).

\bibitem{knysh14}
S.~I. {Knysh}, E.~H. {Chen}, and G.~A. {Durkin}, \enquote{{True Limits to
  Precision via Unique Quantum Probe},} ArXiv e-prints  (2014).

\bibitem{lidke}
K.~Lidke, B.~Rieger, T.~Jovin, and R.~Heintzmann, \enquote{Superresolution by
  localization of quantum dots using blinking statistics,} Opt. Express
  \textbf{13}, 7052--7062 (2005).

\end{thebibliography}

\end{document}